\documentclass{article}\usepackage{jkas2}
\runningauthor{R.Cassano, G.Brunetti, G. Setti}
\runningtitle{Occurence and Luminosity Functions of Giant Radio Halos from 
Magneto-Turbulent Model}
\beginpage{1}
\endpage{5}

\begin{document}

\font\twelvei = cmmi10 scaled\magstep1 
       \font\teni = cmmi10 \font\seveni = cmmi7
\font\mbf = cmmib10 scaled\magstep1
       \font\mbfs = cmmib10 \font\mbfss = cmmib10 scaled 833
\font\msybf = cmbsy10 scaled\magstep1
       \font\msybfs = cmbsy10 \font\msybfss = cmbsy10 scaled 833
\textfont1 = \twelvei
       \scriptfont1 = \twelvei \scriptscriptfont1 = \teni
       \def\mit{\fam1 }
\textfont9 = \mbf
       \scriptfont9 = \mbfs \scriptscriptfont9 = \mbfss
       \def\bmit{\fam9 }
\textfont10 = \msybf
       \scriptfont10 = \msybfs \scriptscriptfont10 = \msybfss
       \def\bmsy{\fam10 }

\def\etal{{\it et al.~}}
\def\eg{{\it e.g.,~}}
\def\ie{{\it i.e.,~}}
\def\lsim{\raise0.3ex\hbox{$<$}\kern-0.75em{\lower0.65ex\hbox{$\sim$}}}
\def\gsim{\raise0.3ex\hbox{$>$}\kern-0.75em{\lower0.65ex\hbox{$\sim$}}}

\title{Occurence and Luminosity Functions of Giant Radio Halos from 
Magneto-Turbulent Model}

\author{R. Cassano$^{1,2}$, G. Brunetti$^{2}$, G. Setti$^{1,2}$}
\address{$^{1}$ Dipartimento di Astronomia ,Univ. di Bologna, Italy}
\address{$^{2}$ Istituto di Radioastronomia del CNR ,Bologna, Italy}

%



\abstract{
We calculate the probability to form giant radio halos ($\sim$1 Mpc size)
 as a function of the mass of the host clusters by using a Statistical 
Magneto-Turbulent Model (Cassano \& Brunetti, these proceedings). 
We show that the expectations of this model are in good agreement with the 
observations for viable values of the parameters. In particular, the abrupt 
increase of the probability to find radio halos in the more massive galaxy 
clusters ($M \gsim 2\times 10^{15}M_{\odot}$) can be well reproduced. 
We calculate the evolution with redshift of such a probability and find that 
giant radio halos can be powered by particle acceleration due to MHD turbulence 
up to z$\sim$0.5 in a $\Lambda CDM$ cosmology. 
Finally, we calculate the expected Luminosity Functions of radio halos (RHLFs). 
At variance with previous studies, the shape of our RHLFs is characterized 
by the presence of a cut-off at low synchrotron powers which reflects 
the inefficiency of particle acceleration in the case of less massive galaxy 
clusters.}


\maketitle

\section {Introduction}

Radio observations of galaxy clusters indicate that the detection rate of
radio halos (RHs) shows an abrupt increase with increasing the X-ray luminosity 
of the host clusters. In particular about 30-35\% of the galaxy 
clusters with X-ray luminosity larger than $10^{45}$ erg/s  show diffuse 
non-thermal radio emission (Giovannini \& Feretti 2002); these clusters 
have also high temperature (kT $>$ 7 keV) and large mass 
($\gsim$ 2$\times$ $10^{15} M_{\odot}$). 
Furthermore giant RHs are frequently found in merging clusters 
(e.g., Schuecker et al 2001). 
These observations suggest that there is a connection between 
thermal and non-thermal phenomena in galaxy clusters.\\
Recent papers (Ensslin and R$\ddot{o}$ttgering 2003; Kuo et al. 2004) have 
investigated the statistics of RHs and their connection with the thermal 
properties of the host clusters from a theoretical point of view. 
These works are based on assumpions in defining the condition of RHs 
formation from observational correlations and/or 
mass thresholds. Present data suggest that giant RHs may 
be accounted for by synchrotron emission from relativistic electrons 
reaccelerated by the turbulence generated in the cluster volume during 
merger events (Brunetti 2003; Brunetti this proceedings). 
Thus, with the aim to investigate the statistical properties and the 
connection between thermal and non-thermal phenomena in galaxy clusters, we have 
developed a statistical magneto-turbulent model (Cassano \& Brunetti 2004, 
{\bf C\&B model}; Cassano \& Brunetti these proceedings) in which we follow the 
formation of clusters of galaxies (making use of the extended Press \& Schechter (1974) 
formalism) and estimate the injection of fluid turbulence and of fast magnetosonic (MS) waves 
during cluster mergers. Then we calculate the evolution of the electron spectra in the ICM  and the 
resulting radio (synchrotron) and hard X-ray (Inverse Compton) emission spectra. 
By using our model we can thus investigate the probability of 
formation of RHs in a well defined physical framework, the evolution with 
redshift of such a probability and the expected Luminosity Functions of RHs (RHLFs).
Here we applay the C\&B model under the following assumptions:

\begin{itemize}
\item[-] We focus the attention on the formation of giant RHs only 
(radius $R_{H}\sim 500$ $h_{50}^{-1}$ Kpc).
\item[-] The magnetic field strength averaged over the emitting volume is assumed 
to be $<B>\sim 0.5$ $\mu$G, independent on the mass of the parent cluster.
\end{itemize}
The adopted cosmologies are: EdS ($H_{o}=50$ Km $s^{-1}$$Mpc^{-1}$, $\Omega_{o,m}=1$) and 
$\Lambda$CDM ($H_{o}=70$ Km $s^{-1}$$Mpc^{-1}$, $\Omega_{o,m}=0.3$, $\Omega_{\Lambda}=0.7$, $\sigma_8=0.9$).

\section {Occurence of RHs: predictions vs observations}


By making use of the C\&B model we have run Monte Carlo simulations to obtain a sufficiently 
large number of merger trees in order to have a large synthetic population of galaxy 
clusters with a wide range of present day masses and temperatures. 
In this way we are able to statistically follow the cosmological evolution of the non-thermal emission 
and of the properties of the thermal ICM. Clusters with RHs in our synthetic population 
are identified with those objects with a synchrotron cut-off 
$\nu_{b}\gsim 10^{2}$ MHz in a region of 1 $Mpc$ $h_{50}^{-1}$ size. 
We have calculated the probability to form RHs ($z\leq 0.2$) 
in two mass bins: binA=$[1.8-3.6]\times10^{15}M_{\odot}$ and binB=$[0.9-1.8]\times
10^{15}M_{\odot}$ (EdS cosmology). These mass bins are consistent with those considerated in 
the observational studies and thus allow us to compare our expectations 
with observations.
In Fig.(1) we report the probability to form a giant ($\simeq$1 Mpc 
$h_{50}^{-1}$ size) RH (red points) in the two mass-bins 
(including the statistical error estimated from our Montecarlo simulations) 
as a function of the parameter $\eta_{t}$, which gives the fraction of energy 
of the turbulent motions injected by cluster merger which is 
channeled in the form of MS waves in the C\&B model. 

\begin {figure}[t]
\vskip 0cm
\centerline{\epsfysize=9.0cm\epsfbox{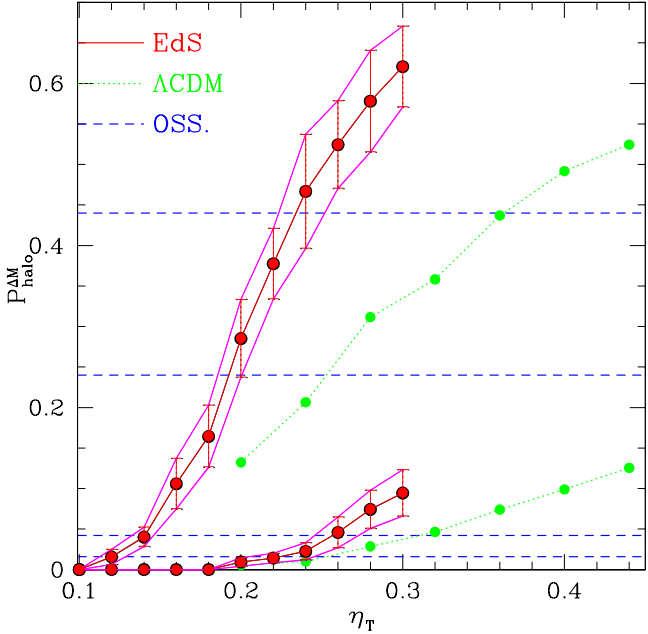}}
\vskip -0.3cm
\label{fig1}
\caption{Expected formation probability of RHs
($R_{H}\simeq 500 h_{50}^{-1}$kpc, $B \sim 0.5 \mu$G)
as a function of parameter $\eta_t$ in a EdS cosmology
(solid lines with error bars) and in a $\Lambda$CDM cosmology
(dotted lines) in the mass bins:
binA=$[1.8-3.6]\,10^{15}M_{\odot}\,h_{50}^{-1}$ and
binB=$[0.9-1.8]\,10^{15}M_{\odot}\,h_{50}^{-1}$ for EdS case
and  binA=$[1.9 - 3.8] \cdot 10^{15}$ M$_{\odot}$h$_{70}^{-1}$ and
binB=$[0.945 - 1.9] \cdot 10^{15}$ M$_{\odot}$h$_{70}^{-1}$
for the $\Lambda$CDM model.
The two bottom dashed lines mark the observed
probabilities for RHs in the mass binB while the
two top dashed lines mark the observed probabilities
in the mass binA; observational regions already account for 
1$\sigma$ errors.}
\vskip -0.2cm
\end{figure}

The dashed blue lines mark the range of the observed probabilities 
(Giovannini et al. 1999) in the binA (top dashed region) and in the binB 
(bottom dashed region), respectively.
For a comparison in Fig.(1) we also report the probability to form a RH 
($z\leq 0.2$) in a $\Lambda$CDM cosmology (green dotted lines). As expected, we
find that at $z\leq 0.2$ the results are relatively indipendent from the 
considered cosmology, with the $\Lambda$CDM model being slightly less efficient.
The main result is that in both EdS and $\Lambda$CDM models it is possible to find a unique
interval of $\eta_{t}$ in which the model reproduces the observed probability
for both the cluster-mass bins. In particular, in agreement with observations and 
indipendently from the adopted cosmology, we find that  20-30\% of clusters 
in the binA can form a RH and that only 2-3\% of galaxy clusters 
in the binB host a RH.
Given the requested values of $\eta_{t}$ (Fig.(1)), we find that the relatively high 
occurence of RHs observed in massive clusters can be well reproduced by our 
particle acceleration model under reasonable conditions, i.e. that a fraction of 
20-30\% of the energy of the turbulent motions injected during cluster merger  
(which corresponds to a few percent of the thermal energy) is
in the form of MS waves.

\begin {figure}[t]
\vskip  0.5cm
\centerline{\epsfysize=10.2cm\epsfbox{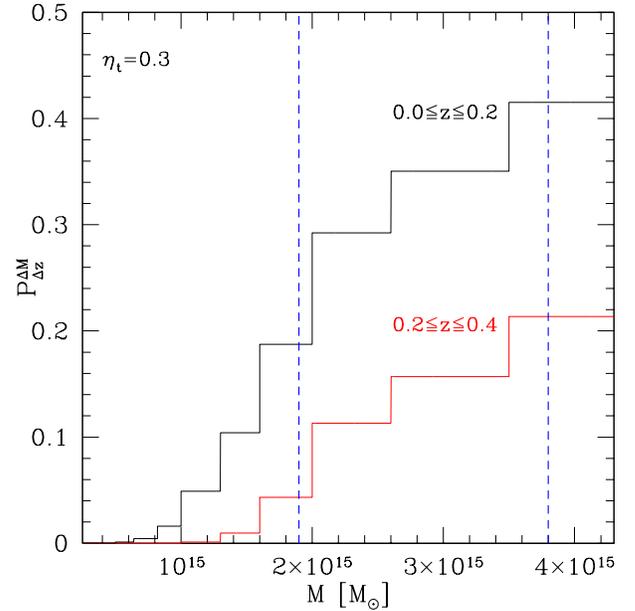}}
\vskip -2cm
\label{fig3}
\caption{Probability to form giant RHs ($> 1 Mpc$ $h_{50}^{-1}$ size) as a
function of the cluster mass in two relevant redshift bins: 
z=0-0.2 (red lines) and z=0.2-0.4 (black lines) in a $\Lambda$CDM model. 
Vertical dashed lines mark the $[1.9-3.8]\times10^{15}M_{\odot}$ mass bin of
Fig.(1). Calculations are obtained for $\eta_t=0.3$.}
\vskip -0.2cm
\end{figure}

\section {Radio Halos Statistics and cluster mass}

In the previous Section we have found that the probabilty 
to form a RH has a strong dependence
on the mass of the host cluster, it goes from few \%
for $M<1.9 \times 10^{15} M_{\odot}$ to 20-30\% for 
$M\gsim2\times 10^{15} M_{\odot}$, 
in agreement with observations (Fig.(1)).

Now we calculate the probability to find RHs as 
a function of the cluster mass from our synthetic population 
of galaxy clusters in a $\Lambda$CDM cosmology: 
this is a crucial point of our model and marks the difference with
previous studies (e.g., Ensslin \& R$\ddot{o}$ttgering (2002)).
As an example, in Fig.(2) we report this probability
in two redshift bins z=0-0.2 (black lines) and z=0.2-0.4 (red lines).
Clusters with mass $<< 10^{15}M_{\odot}$ have a negligible probability 
to form giant RHs; on the other hand, such a probability is found to reach
$\sim$40\% for clusters with masses $ >3\times10^{15}M_{\odot}$ in 
the 0-0.2 redshift bin. For a comparison with Fig.(1) we also mark 
in Fig.(2) the mass bin $[1.9-3.8]\times10^{15}M_{\odot}$ (vertical blue dashed lines).

\begin {figure}[t]
\vskip  0.0cm
\centerline{\epsfysize=10.2cm\epsfbox{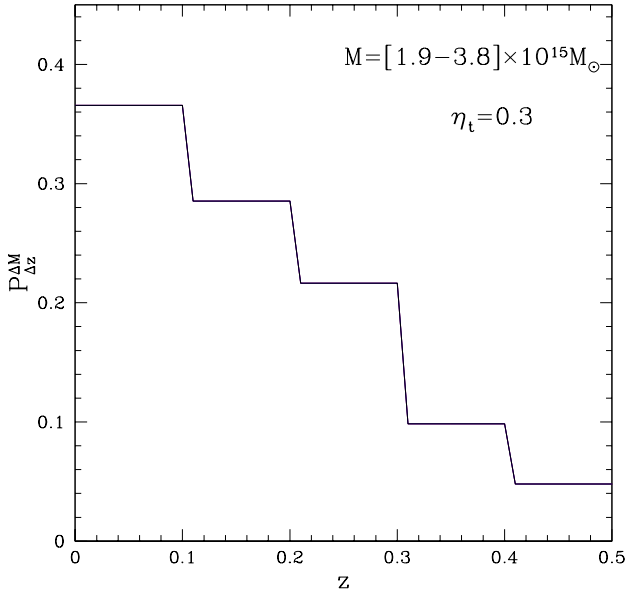}}
\vskip -2.0cm
\label{fig2}
\caption{Probability to form giant RHs ($\sim 1 Mpc$ $h_{50}^{-1}$ size)
with redshift in the mass bin $[1.9-3.8]\times10^{15}M_{\odot} h_{70}^{-1}$
in a $\Lambda$CDM cosmology. Calculations are performed for $\eta_t=0.3$.}
\vskip -0.2cm
\end{figure}

Thus, the important finding of these calculations is that only massive
clusters can host giant RHs ($R_H \geq 500$ kpc $h_{50}^{-1}$) and
that the probabilty to form these diffuse radio sources presents
an abrupt increase for clusters with about
$M \gsim 2 \times 10^{15} M_{\odot}$ (Fig.(1) and Fig.(2)).
These findings can be simply explained in the framework of our model.
Infact, it can be shown that in the C\&B model the energy of the 
turbulence injected in galaxy clusters is expected to 
roughly scale with the thermal energy of the clusters
(Cassano \& Brunetti 2004; Cassano \& Brunetti these proceeding
Fig(2)). This seems reasonable and immediately
implies that the energy density of the turbulence is an
increasing function of the mass of the clusters, i.e. 
${\cal E}_t \propto T \propto M^{a}$ (a=0.56-0.67), and thus particle
acceleration is favoured in massive clusters.

In general, the infall of subclusters through a main cluster,
which is not very massive, injects turbulence in a volume $V_t$ 
(calculated using {\it Ram Pressure Stripping}, Fujita, Takizawa, Sarazin 2003; 
Cassano \& Brunetti 2004) which is found to be smaller than that of giant RHs, 
$V_H$ ($V_H=4\pi R_H^3/3$), and thus the efficiency of the 
mechanism is reduced by about a factor of $V_t/V_H$ 
in the case of less massive clusters. 
On the other hand, major mergers between massive subclusters
are expected to inject turbulence on larger volumes,
of the order of $V_H$, and thus the efficiency of the 
generation of RHs is not reduced and this further favour massive objects as the 
parent clusters of RHs.\\

More quantitatively, it can be shown (Cassano \& Brunetti 2004) that the
acceleration efficiency $\chi$ (within $V_H$), triggered by a major merger event 
scales about with $\chi \propto M^{0.75-1.25}$ 
(0.75 for $M \geq 3 \cdot 10^{15}M_{\odot}$, 1.25 for $M < 10^{15}M_{\odot}$). 
Since the maximum energy of the accelerated electrons 
is $\gamma_b\propto\chi/(B^2+B_{CMB}^2)$, where $B_{CMB}=3.2\cdot (1+z)^4 \mu G$ 
is the strength of the equivalent magnetic field of the CMB, 
and the break frequency is $\nu_b\propto\gamma_b^2 B$, one has: 

\begin{equation}
\nu_b\propto M^{1.5-2.5}{{B}\over{(B^2+B_{CMB}^{2})^{2}}}
\label{nub}
\end{equation}

and consequently massive clusters are statistically favourite to have 
$\nu_b\gsim10^2$ MHz (which is the adopted condition to define 
the presence of a RH).

\section {Evolution of radio halos with redshift}

The probability to form RHs dependes on the combination of the energy 
losses suffered by relativistic electrons (mainly due to IC losses 
$\propto(1+z)^{4}$) with the acceleration efficiency powered by the 
turbulence generated during cluster mergers (which depends on the merger history).

\begin {figure}[t]
\vskip 0cm
\centerline{\epsfysize=10.2cm\epsfbox{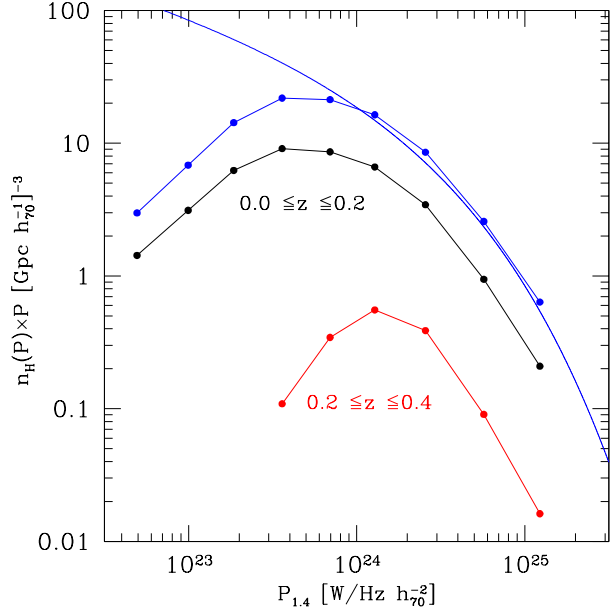}}
\vskip -2cm
\label{fig4}
\caption{RHLFs ($n_{H}(P)\times P$) expected by the C\&B model. 
Results are shown for the redshift bins $0\leq z\leq 0.2$ (black curve with points) 
and $0.2\leq z \leq 0.4$ (red curve with points). The expected Local RHLF 
(blue curve with points) is also reported together with the Local RHLF from 
Ensslin \& R$\ddot{o}$ttgering (2002) (solid blue line) for a comparison.}
\vskip -0.2cm
\end{figure}

In this Section we calculate the evolution with redshift of the 
probability to form RHs in a $\Lambda$CDM cosmology. As an example, 
in Fig.(3) we report the probability to form a giant 
RH ($\simeq$ 700 Kpc in a $\Lambda$CDM cosmology) 
with redshift in the mass bin: $[1.9-3.8]\times10^{15}M_{\odot}$ 
for a representative value of $\eta_t$ ($\eta_t=0.3$).
The occurrence of RHs decreases with redshift due to the higher IC 
energy losses. We note however that such a decrease is not dramatic since 
in a $\Lambda$CDM Universe major mergers develope at slightly higher redshift with 
respect to a EdS Universe. For instance, in the considered case the formation rate of 
RHs is 20-36\% at relatively low redshift and decreases to 10\% 
at higher redhifts ($0.3\leq z\leq 0.4$).\\

\section {The Luminosity Functions of Radio Halos (RHLFs)}

We have alredy shown that the observed probability to find RHs with the cluster 
mass is well reproduced by the C\&B model (see Sec.2). In Cassano \& Brunetti 
2004 (see also Cassano \& Brunetti these proceedings) it has also been shown 
that the typical synchrotron and IC luminosity of RHs can be well reproduced 
by the model assuming that during mergers a few percent of the thermal energy of 
the cluster is in the form of MS waves (\ie $\eta_{t}>0.1-0.2$).
Given these promising results, in this Section we derive the expected
luminosity functions of giant RHs (RHLFs).
First we use the probility to form RHs with the cluster's mass $P_{\Delta M}^{\Delta z}$ 
(Fig.2) to estimate the mass functions of RHs ($dN_{H}(z)/dM dV$):

\begin{equation}
{dN_{H}(z)\over{dM\,dV}}=
{dN_{cl}(z)\over{dM\,dV}}\times P_{\Delta M}^{\Delta z}=n_{PS}\times
P_{\Delta M}^{\Delta z},
\label{RHMF}
\end{equation}

where $n_{PS}=n_{PS}(M,z)$ is the Press \& Schechter (1974) mass function 
(we use $n_{PS}$ since our model is based on Press \& Schechter formalism). 
The RHLF is given by:

\begin{equation}
{dN_{H}(z)\over{dV\,dP_{1.4}}}=
{dN_{H}(z)\over{dM\,dV}}\bigg/ {dP_{1.4}\over dM}.
\label{RHLF}
\end{equation}

In order to derive $dP_{1.4}/dM$ in Eq.(\ref{RHLF}), we combine the observed 
correlations between radio power at 1.4 Ghz  ($P_{1.4 GHz}$) and bolometric 
X-ray luminosity ($L_X$) (e.g., Feretti 2003) and between $L_{X}$ and the 
virial mass, $M_{200}$ (e.g., Arnaud \& Evrard 1999). 
The used $P_{1.4 GHz}$-M correlation is obtained collecting the data from 
all the known clusters with giant RHs and converting them in 
a $\Lambda$CDM cosmology. In Fig.(4) we report the Local (here calculated
for $z<0.05$) RHLF (number of RHs per $Gpc^3$ as a function of the radio power) expected by our model and the expected RHLFs 
in the bins $0\leq z\leq 0.2$ and  $0.2\leq z \leq 0.4$ (lines with points). 
Our RHLFs are compared with the Local $(RHLF)_{E\&R}$ (blue solid line) 
of Ensslin \& R$\ddot{o}$ttgering (2002). The $(RHLF)_{E\&R}$ is obtained combining 
the X-ray observed luminosity function of clusters with the radio luminosity - 
X-ray luminosity correlation and assuming that a costant fraction $f_{rh}=1/3$ 
of galaxy clusters have RHs. 
The most important difference between the two luminosity functions 
is that our RHLF shows a cut-off/flattening at low radio powers.
We stress that the flattening at low powers is a unique feature of particle acceleration
models since it marks the effect of the decrease of the efficiency of the 
particles acceleration in the case of the less massive galaxy clusters and consequently 
the presence of a synchrotron cut-off $\nu_{b}< 10^{2}$ MHz. \\
Future radio observations (e.g., with LOFAR and LWA) should be able to 
test the presence of such a low-power cut-off in the RHLFs and the evolution
of the RHLFs with redshift.

\acknowledgements{G.B. and R.C. acknowledge partial support from CNR grant
CNRG00CF0A.}

\end{document}